\documentclass[12pt]{iopart}
\usepackage{iopams,epsfig}

\def\d{\delta}

\def\k{\kappa}
\def\vt{\vartheta}
\def\ep{\epsilon}
\def\ep{\epsilon}
\def\L{{\mathcal{L}}}
\def\E{{\mathcal{E}}}

\def\C{{\mathcal{C}}}
\def\F{{\mathcal{F}}}
\def\S{{\mathcal{S}}}
\eqnobysec
\begin{document} 
\title{Noether currents and charges for Maxwell-like Lagrangians}
\author{Yakov Itin$^{1,2}$}
\address{$^{1}$Institute of Mathematics,  
Hebrew University of Jerusalem,  91904 Jerusalem, Israel\\
$^{2}$Jerusalem College of Engineering,
  91035 Jerusalem, Israel\\ 
{\tt itin@math.huji.ac.il}}

\begin{abstract} 
Hilbert-Noether theorem states that a current associated to diffeomorphism invariance 
 of a Lagrangian 
vanishes on shell modulo a divergence of an arbitrary superpotential. 
Application of the Noether procedure to physical Lagrangians yields, 
however, meaningful (and measurable) currents. 
The well known solution to this ``paradox'' is to involve the variation of the 
metric tensor. Such procedure, for the field considered on a 
fixed (flat) background, is sophisticated logically (one need to 
introduce the variation of a fixed field) and formal.   
 
We analyze the Noether procedure for a generic diffeomorphism invariant 
$p$-form field model.  
We show that the Noether current of the field considered on a variable background 
coincides with the current treated in a fixed geometry. 
Consistent description of the canonical 
energy-momentum current is possible only  if the dynamics of the 
geometry (gravitation) is taken into account. 
However, even the ``truncated'' consideration yields the proper expression. 
We examine the examples of  the  free $p$-form gauge field theory, 
the GR in the coframe representation and the metric-free electrodynamics. 
Although, the variation of a metric tensor is not acceptable in the latter case, 
the Noether procedure yields the proper result.
\end{abstract} 

\pacno{04.20.-q, 11.15.-q, 11.30.-j}

\section{Introduction}                  
Probably, the main problem of the Noether procedure 
is to establish a reasonable correspondence between the set of 
the physical meaningful and measurable currents on  the one hand 
and the set of the formal Noether currents on the other hand 
\cite{Julia:1998ys}-\cite{Hehl:ue}. 
It is well known that the relation between these two types of quantities 
is highly non-trivial. In particular, the  Noether theorem states that 
a current associated to a gauge symmetry of a Lagrangian 
necessarily vanishes on shell modulo a divergence of 
an arbitrary superpotential. 
Consequently, this Noether current is not observable and physically 
meaningless. On the other hand the superpotential 
(Noether charge) is known to play the crucial role in Wald's analysis 
of the black hole entropy \cite{Wald:nt}, \cite{Iyer:1994ys} .   

It was recognized long ago that, in order to have a proper description of 
the electromagnetic current, one has to extend the pure electrodynamic 
system by introducing  charged scalar and/or fermionic fields. 

We will show in this paper that the situation for the energy-momentum current 
is fairly similar. 
Namely, the consistent description of the energy-momentum current via 
the Noether procedure is only possible if the gravitational 
field is taken into account.  

The purpose of this paper is to carry out this task for 
a $p$-form gauge field $\psi$, or, 
 as one also says, for  an antisymmetric tensor field \cite{Henneaux:1996ws}, 
\cite{Freedman:us}. 
This field can be furnished with  some interior indices, 
i.e., it can be  tensorial valued. 
We assume that the Lagrangian depends only on the field  
$\psi$ and its exterior derivative $d\psi$. 
Certainly, our requirements are rather restrictive and take out of the 
consideration a wide class of mathematically interesting models. 
The aim of this paper is to examine how the problems of the Noether  
procedure appear in this relatively simple context.
Note, that even with these restrictions, our framework is suitable for the description of   
many interesting physical models, including gravity.  

The organization of the paper is as following:   

In the next section, neglecting  any geometric features of the manifold, 
we recall the standard Noether procedure for a generic Lagrangian 
of a $p$-form field. 
We restate the well known fact that the Noether current, associated with 
diffeomorphism invariance,  vanishes identically on shell. 
It should be noted that our treatment is completely local, so diffeomorphism 
invariance always refers to  local diffeomorphism transformations. 
In particular, we  ignore the topological obstructions  (such as non-orientability) 
which can forbid the existence of global frame fields and of global volume forms.

 We argue that this triviality of the conserved current appears 
because the geometry of the manifold is completely ignored in 
this consideration. 
A viable Lagrangian density has to be represented by an odd (twisted) 
differential form. 
Such Lagrangian   cannot be constructed  only out of  even forms: 
the field and its derivative. 
It has always to  involve some odd operator  of the  type of 
Hodge's dual map, which changes the parity of the form.  
Such odd operator necessary depends on  the geometric 
features of the manifold, similarly to the Hodge map, which depends on the 
metric tensor.     
 
In order to find out a non-trivial conserved current, we 
consider in the second section a model for a  $p$-form field given 
on a fixed coframe background. 
The variation procedure  takes now into account, with some restrictions, 
the variation of the coframe field in addition to the variation of the 
field $\psi$. 
The total Noether current of the system vanishes on shell. 
However, now we are able to identify a non-trivial energy-momentum 
current for the  field $\psi$ with a piece of this trivial 
total Noether current. 
The derivation involves, however, some logical contradictoriness.  
Namely, the coframe field has to be  considered 
as  subject to variations, even it is fixed.   
The  situation is similar to the derivation of the Hilbert energy-momentum 
tensor for a field given on a fixed manifold. 
This justifies the necessity to involve a dynamical coframe field, or, 
in physical language, to consider the gravity field together with the 
field $\psi$. 

In the third section we consider a system of a $p$-form field and 
a coframe field, both dynamical. 
We show that the derivative of the Lagrangian with respect to the coframe 
(Hilbert current) plays a role of the conserved source in the coframe 
field equation. 
Again, the total Noether current, associated with the  
diffeomorphism invariance  of the system, vanishes on shell. 
However, this equation represents now a relation between  
the Noether and the Hilbert currents. Moreover, we derive in this way    
an explicit expression of the Hilbert current, of the Noether charge, and 
the Noether identity. 
All these quantities are well defined.

In the fourth section we deal with a generic  first order Lagrangian shifted 
by a total derivative. 
We show that such shift preserves the field equations as well as 
the conserved energy-momentum current. The Noether charge is shifted, however. 

In the fifth section we consider three examples of generalized 
Maxwell Lagrangians. 
We examine the  free $p$-form gauge field theory, 
the GR in the coframe representation and the metric-free electrodynamics.  
\section{A $p$-form Lagrangian}
\subsection{Non-geometric Lagrangian}
Consider a $p$-form field $\psi$ defined on an $n$-dimensional 
differential manifold $M$. 
It is a straightforward generalization of the ordinary  1-form potential 
field $A$ of $4D$ Maxwell electrodynamics.
We describe the dynamics of the field $\psi$ 
by a generic Lagrangian $n$-form of first order, i.e., 
only the field and its exterior (first order) derivative are involved:  
\begin{equation}\label{form-1}
\L=\L(\psi,d\psi)\,,
\end{equation}
see for instance  \cite{Julia:1998ys}. 
Denote the  derivatives of the Lagrangian 
taken with respect  to the field $\psi$ by 
\begin{equation}\label{form-2}
\sigma:=\frac{\partial \L}{\partial\psi}\,, \qquad 
\pi :=\frac{\partial \L}{\partial(d\psi)}\,.
\end{equation}
We will refer to the   $(n-p)$-form    $\sigma$ as the 
{\it  current of the field} $\psi$, and to the   $(n-p-1)$-form $\pi$ as 
the {\it field strength}.  

Using these abbreviations,  
the variation of the Lagrangian (\ref{form-1}), 
in the exterior form notations, may be written  as 
\begin{equation}\label{form-3}
\d \L=\d\psi\wedge\sigma+\d (d\psi)\wedge \pi\,.
\end{equation} 
Applying the commutativity of the operators $d$ and $\d$,  
we extract the total derivative and obtain the variational relation 
\begin{equation}\label{form-4}
\d \L=\d\psi\wedge\E+d\Omega\,,
\end{equation}
where the  $(n-p)$-form 
\begin{equation}\label{form-5}
\E:=\sigma-(-1)^pd\pi\, 
\end{equation}
is the action of the Euler-Lagrange operator on the field $\psi$.  
The  $(n-1)$-form $\Omega$ is   defined as  
\begin{equation}\label{form-6}
\Omega:=\d\psi\wedge \pi\,.
\end{equation}
The  form $\Omega$ is  linear in variations of the field. 
This quantity is sometimes referred to as 
the {\it pre-symplectic potential} \cite{Wald:nt,Iyer:1994ys}. 
Observe that for a given Lagrangian both quantities 
$\E$ and $\Omega$ are well-defined without any ambiguity. 

Consider the variations of the fields which vanish 
at at a boundary of a region. 
We obtain the  field equation which immediately yields the conservation 
law for the $(n-p)$-form $\sigma\,$:
\begin{equation}\label{form-7}
\E=0 \quad {\rm or}\quad d\pi=(-1)^p\sigma\,\quad\Longrightarrow\quad  d\sigma=0\,.
\end{equation}
Returning to the variational relation (\ref{form-4}), we recognize 
a special case, when  the variation of the Lagrangian is an exact form 
$\d \L=dS$. Eq.(\ref{form-4})  implies now the existence of  an $(n-1)$-form 
$\Theta:=-S+\Omega\,$
which is conserved modulo the field equations, i.e., on shell: 
\begin{equation}\label{form-8}
d \Theta+\d\psi\wedge\E=0\,\quad\Longleftrightarrow \quad d\Theta\approx 0\,.
\end{equation}
Here and in the sequel, we use the symbol $\approx$ for 
``equal up to a linear combination of the field equation form $\E$''. 

The Lagrangian (\ref{form-1}) depends 
only on the exterior form field and its exterior derivative,  
thus it is  diffeomorphism invariant.
In order to specialize the  diffeomorphism invariance,   
the variations of the field have to be  generated by the
Lie derivative taken with respect to a smooth vector field $\xi $, 
\begin{equation}\label{form-9}
\d \psi=L_\xi\psi=d(\xi\rfloor\psi)+\xi\rfloor d\psi\,.
\end{equation}
Because of the  diffeomorphism invariance,
 the  variation of the $n$-form Lagrangian is
induced by the Lie derivative taken with respect to the same vector
field $\xi$, i.e.,
\begin{equation}\label{form-10}
\d \L=L_\xi\psi=d(\xi\rfloor \L)\,.
\end{equation}
Accordingly,  we have an   $(n-1)$-form  
\begin{equation}\label{form-11}
\Theta(\xi)=-\xi\rfloor \L+\Omega(\xi)\,, 
\end{equation}
generated by the diffeomorphism symmetry of the Lagrangian,  
which is weak (on shell) conserved 
\begin{equation}\label{form-11*}
d\Theta(\xi)+\d\psi\wedge\E=0\,, \qquad d\Theta(\xi)\approx 0\,.
\end{equation}
We will refer to $\Theta(\xi)$ as the {\it Noether current}. 
The explicit expression for this quantity  is derived from 
(\ref{form-9}, \ref{form-10},\ref{form-11}) as 
\begin{equation}\label{form-12}
\Theta(\xi)=-\xi\rfloor \L+\big[(\xi\rfloor d\psi)+
d(\xi\rfloor \psi)\big]\wedge \pi\,.
\end{equation}
Because the Lagrangian depends only on $\psi$ and $d\psi$, 
also the field strength $\pi$ depends only on these variables. 
Thus, the Noether current is locally constructed only out 
of the quantities $\psi, d\psi$,   
and of  the undefined vector field $\xi$.
Extracting the total derivative in (\ref{form-12}) 
and using the field equation 
(\ref{form-7}), we decompose the  Noether current  as 
\begin{equation}\label{form-13}
\Theta(\xi)=\S(\xi)+dQ(\xi)\,,
\end{equation}
where 
\begin{equation}\label{form-14}
\S(\xi):=-\xi\rfloor \L+(\xi\rfloor d\psi)\wedge\pi+
(\xi\rfloor \psi)\wedge\sigma \,
\end{equation}
is an $(n-1)$-form current, whereas  
\begin{equation}\label{form-15}
Q(\xi):=(\xi\rfloor \psi)\wedge\pi \,
\end{equation}
is an $(n-2)$-form  charge. Certainly, two currents are conserved on shell 
simultaneously,    
\begin{equation}\label{form-16}
d\Theta(\xi)\approx 0\quad\Longleftrightarrow \quad d\S(\xi)\approx 0\,
\end{equation}
for an arbitrary vector field $\xi$. 
 
As it is proved in \cite{Iyer:1995kg},   
a decomposition of a type (\ref{form-13}) may be provided for 
an arbitrary diffeomorphism invariant Lagrangian.
It means that a total derivative can be extracted from   
the Noether current in such a way that the remaining term $\S(\xi)$ 
is (algebraically) linear in the undetermined vector field $\xi$. 
Decompose $\xi$ into its components according to $\xi=\xi^ae_a$. 
The current $\S(\xi)$ 
involves the undefined vector field $\xi$ only in a linear algebraic form.  
Hence it fulfills
\begin{equation}\label{form-16a}
\S(\xi)=\xi^a\S(e_a)\,. 
\end{equation}
Thus the conservation law for this current reads  
\begin{equation}\label{form-17}
d\S(\xi)=d\xi^a\wedge \S(e_a)+\xi^a d\S(e_a)\approx 0\,.
\end{equation}
The arbitrariness of  $\xi$ means independence of the quantities 
$\xi^a$ and $d\xi^a$. 
Hence, two terms in the   right hand side of (\ref{form-17}) should vanish
simultaneously. Thus we obtain two, so-called, {\it cascade equations}  
\cite{Julia:1998ys}, \cite{Kopczynski:af}
\begin{eqnarray}\label{form-18}
\xi^a&:& \qquad d\S(e_a)\approx 0\,,
\\ \label{form-19} 
d\xi^a&:& \qquad \S(e_a)\approx 0\,.
\end{eqnarray}
Observe that (\ref{form-18}) is not merely a consequence of the equation 
 (\ref{form-19}). Indeed,  (\ref{form-19}) means $\S(e_a)={A_a}^b\wedge \E_b$, 
for some $(p-1)$-form ${A_a}^b$. Thus its exterior derivative is not, 
in general, proportional to the field equation. 

Because of (\ref{form-16a},\ref{form-19}),  the current 
$\S(\xi)$ vanishes on shell, i.e., on all solutions 
of the field equation, for an arbitrary vector field $\xi$ 
\begin{equation}\label{form-20}
\S(\xi)\approx 0\,.
\end{equation}
Inserting (\ref{form-20}) into (\ref{form-13}) we obtain 
\begin{equation}\label{form-21}
\Theta(\xi)\approx dQ(\xi)\,.
\end{equation}
In this way we derive the well known property of  gauge conserved 
currents (Noether-Hilbert theorem). 
The conserved Noether current $\Theta(\xi)$, which  corresponds to the gauge 
invariance of the Lagrangian
(diffeomorphism in our case),   is exact  on shell. 
Although some technical differences, our differential form consideration 
is similar to the tensorial treatment in \cite{Julia:1998ys}. 

Let us take into account the intermediary result (\ref{form-20}) 
of our derivation: 
the vanishing  of the  current $\S(\xi)$ on shell. 
In the case of the ordinary Maxwell field 
with $p=1$ , $\psi=A$, and  
\begin{equation}\label{form-22}
\L=-\frac 12 dA\wedge*dA
\end{equation}
Eq.(\ref{form-14}) takes the form 
\begin{equation}\label{form-23}
\S(e_a)=-\frac 12 \, e_a\rfloor (dA\wedge*dA)+(e_a\rfloor dA)\wedge*dA)\,.
\end{equation}
Thus,  it is completely identical to the electrodynamic 
energy-momentum current. 
This current  uniquely defines the ordinary measurable 
energy-momentum tensor \cite{book}. 
Thus $\S(\xi)$ cannot vanish for the electrodynamic fields identically. 
Consequently, at least in the case of the free Maxwell field,  
we seem to reach a  ``contradiction''  to the 
Noether-Hilbert theorem. 

It is well known, that the problem comes from elimination of the 
geometric variables from the variation procedure. 
The consideration above is, in fact, restricted to Lagrangians which depend 
only on the fields and its derivatives. 
Such Lagrangians are  even (untwisted)  forms, thus  
it is rather naturally, that they do not contribute to the  measurable 
energy-momentum quantities. 
Dicke \cite{dicke} proved that it is almost impossible 
to construct a nontrivial 
Lagrangian for a field "interacted only  with itself\,". 
Only for the metric field a nontrivial Lagrangian can be constructed 
(Hilbert-Einstein). 
All other physical Lagrangians are, in fact, represent interaction 
with some other field (the metric field in most cases). 
This is a situation appeared in (\ref{form-22}),  
where the metric is involved implicitly 
by the Hodge operator, which makes the Lagrangian an odd form.  

\subsection{Non-dynamical coframe}
A non-trivial Lagrangian has to be represented by an 
 odd (twisted) $n$-form. 
We assume the field  $\psi$ to be even (untwisted). 
Thus, in addition to   the even form $\psi$ and its derivatives, 
the Lagrangian has to include some odd operator on forms. 
This odd operator necessarily  inherits certain geometrical 
properties of the manifold. 
It can be, for instance,  the ordinary Hodge dual map or the constitutive 
tensor of the metric-free electrodynamics \cite{book}. 
Such operator may be defined  by the metric tensor or by the coframe field. 
For the time being, we do not specify the odd operator used and 
only assume that the Lagrangian depends also on a fixed coframe field $\vt^a$ 
\begin{equation}\label{nd-1}
\L=\L(\psi,d\psi,\vt^a)\,.
\end{equation}  
Certainly, we require  the coframe field to be essentially 
involved in the $p$-form field Lagrangian. 
 It means that the derivatives 
$ {\partial \L}/{\partial \vt^a}$ and $ {\partial \L}/{\partial (d\psi)}$ are 
assumed to be non-zero functions of $d\psi$ and $\vt^a$. 
 
The variation of this Lagrangian, taken with 
respect to $\psi$ and $\vt^a$, is 
\begin{eqnarray}\label{nd-3}
\d \L&=&\d\psi\wedge\sigma+\d d(\psi)\wedge \pi+
\d\vt^a\wedge\frac{\partial \L}{\partial\vt^a}\nonumber\\
&=&\d\psi\wedge\E+\d\vt^a\wedge\frac{\partial \L}{\partial\vt^a}+
d(\d\psi\wedge \pi)\,.
\end{eqnarray}

In absence of constrains, the general variation procedure requires 
to consider independent variations of all 
fields involved in the Lagrangian. 
Thus in addition to the field equation 
\begin{equation}\label{nd-4}
\E=0 \quad\Longleftrightarrow \quad d\pi=(-1)^p\sigma\,
\end{equation}
we obtain  ${\partial \L}/{\partial\vt^a}=0$. 
It means that  the field $\vt^a$ can not be incorporated into the Lagrangian,  
at all.  
In order to overcome this obstacle  we 
require only the variation $\d\psi$ to be free.  
It means that  the coframe field is considered to be non-dynamical. 
Thus we have only one field equation (\ref{nd-4}).
The variation of the Lagrangian on shell remains in  the form
\begin{equation}\label{nd-5}
\d \L\approx \d\vt^a\wedge\frac{\partial \L}{\partial\vt^a}
+d(\d\psi\wedge \pi)\,.
\end{equation}
Observe that, in contrast to (\ref{form-4}), 
the right hand side of (\ref{nd-5}) is not a total derivative. 
We may apply, however, the diffeomorphism invariance of  the Lagrangian 
also in this case. 
Consider again the variation of the fields to be 
produced by the Lie derivatives. 
In accordance with the non-dynamical nature of the coframe field, we will 
require $d\vt^a=0$. 
Thus the relation (\ref{nd-5}) takes the form 
\begin{equation}\label{nd-7}
d\Theta(\xi)+ d(\xi\rfloor \vt^a)\wedge\frac{\partial \L}
{\partial\vt^a}\approx 0\,,
\end{equation}
where $\Theta(\xi)$ is defined in (\ref{form-12}).
We use (\ref{form-13}) to obtain 
\begin{equation}\label{nd-7*}
d\S(\xi)+ d(\xi\rfloor \vt^a)\wedge\frac{\partial \L}
{\partial\vt^a}\approx 0\,,
\end{equation}
where $\S(\xi)$ is defined in (\ref{form-14}).
This equation has to be satisfied for an arbitrary vector field $\xi$. 
Spelling out (\ref{nd-7}) explicitly  for $\xi=\xi^ae_a$, we obtain  
\begin{equation}\label{nd-8}
d\xi^a\wedge\Big[\S(e_a)+\frac{\partial \L}{\partial\vt^a}\Big]
+\xi^a d\S(e_a)\approx 0\,.
\end{equation}
Independence of the quantities $\xi^a$ and $d\xi^a$ yields two
 cascade equations 
\begin{eqnarray}\label{nd-9}
 \xi^a&:& \qquad d\S(e_a)\approx 0\,,
\\ \label{nd-10} 
d\xi^a&:& \qquad \S(e_a)\approx -\frac{\partial \L}{\partial\vt^a}\,.
\end{eqnarray}
Thus we resolve the contradiction mentioned above.  
The first cascade equation represents the weak conservation law 
for the current $\S(e_a)$. This current does not vanish now, 
in contrast to  (\ref{form-19}). 

In the tensorial approach of field theory, the derivative 
$\partial \L / \partial g_{\mu\nu}$ represents the  Hilbert energy-momentum 
tensor. 
In the coframe approach the similar meaning may be given to the derivative 
$\partial \L / \partial \vt^a$. 
Consequently the second cascade equation represents the equality between the 
{\it canonical current} $\S(e_a)$ and the {\it coframe Hilbert current} 
${\partial \L}/{\partial\vt^a}$.

The price of this result is some non-completeness of the variation procedure. 
(i) We were forced  to consider the coframe field as fixed and non-dynamical, 
however we have to take the variation of the Lagrangian also 
with respect to this field.  
(ii) The condition $d\vt^a=0$ was applied, thus we restricted ourself 
to consider only  holonomic coframes.

In order to resolve these problems, we have to make the 
coframe field dynamical. 

\section{Matter-coframe system}
\subsection{Lagrangian and field equations}
Let be given an $n$-dimensional, smooth, orientable, differential manifold 
$M$. 
We describe the geometry on $M$ by a smooth coframe field 
$\vt^a$ and its dual: a frame field $e_a$, where $a=1,\cdots,n$. 
The duality is expressed by the relation $e_a\rfloor \vt^b=\d_a^b$, where 
$\rfloor$ is the interior product operator. 
The coframe field $\vt^a$ is a set of $n$ even (untwisted) 1-forms, 
which are linear independent at every point of $M$. 
The duality relation provides the linear independence also of  
the frame  field $e_a$ (a set of $n$ even vector fields).
In the case of the teleparallel (coframe) approach to gravity,  
the manifold $M$ is endowed also with a metric 
$g=\eta_{ab}\vt^a\otimes\vt^b$. 
We will not use time being the metric tensor, thus we are 
working on a metric-free background endowed with a coframe field. 

Assume the matter to be represented by a $p$-form field $\psi$. 
Certainly a viable matter system has to include also some set of fermionic 
fields, which we exclude from the consideration, for the sake of simplicity.
Thus we are dealing with some generalization of the Maxwell-Einstein system.  

We assume the fields $\vt^a, e_a$ and $\psi$ to be even (untwisted). 
It means that they are invariant  under a change of orientation of the 
manifold.
We describe the matter-coframe system $\{\vt^a,\psi\}$ 
by a generic Lagrangian form of first order:   
\begin{equation}\label{mat-1}
\L=\L(\psi,d\psi,\vt^a,d\vt^a)\,.
\end{equation}
A non-trivial Lagrangian has to be represented by an 
 odd (twisted) $n$-form. 
Thus, in addition to the even forms $\psi$, $\vt^a$ and their derivatives, 
the Lagrangian has to involve some odd operator on forms. 
We do not specify the odd operator and
only assume that this operator can be expressed in terms
of the coframe field.
Denote the derivatives taken with respect to the coframe field as 
\begin{equation}\label{mat-3}
\Sigma_a:=\frac{\partial \L}{\partial\vt^a}\,, \qquad 
\Pi_a :=\frac{\partial \L}{\partial(d\vt^a)}\,.
\end{equation} 
The odd $(n-1)$-form $\Sigma_a$ will be  referred to as the 
{\it current of the coframe field} 
while the  odd $(n-2)$-form $\Pi_a$ as the {\it strength of the coframe field}.
The derivatives of the Lagrangian taken with respect to the matter 
field are defined in (\ref{form-2}). 
We will refer now to the  odd $(n-p)$-form    $\sigma$  as the 
{\it current of the matter field}, and to the  odd $(n-p-1)$-form $\pi$ as 
the {\it strength of the matter field}. 

Using the abbreviations (\ref{form-2}) and (\ref{mat-3}), 
variation of the Lagrangian may be written as 
\begin{equation}\label{mat-4}
\d \L=\d\psi\wedge\sigma+\d (d\psi)\wedge \pi+\d\vt^a\wedge\Sigma_a+
\d (d\vt^a)\wedge \Pi_a\,.
\end{equation} 
Extracting the total derivatives, we obtain the variational relation 
\begin{equation}\label{mat-5}
\d \L=\d\psi\wedge\,^{\rm (mat)}\E+\d\vt^a\wedge\,^{\rm (gr)}\E_a+d\Omega\,,
\end{equation}
where the field equation forms are 
\begin{eqnarray} \label{mat-5a}
\,^{\rm (mat)}\E&:=&\sigma+(-1)^{p+1}d\pi\,,\\
\,^{\rm (gr)}\E_a&:=&\Sigma_a+d\Pi_a\,,\label{mat-5b}
\end{eqnarray}
while the pre-symplectic potential is 
\begin{equation}\label{mat-5c}
\Omega:=\d\psi\wedge \pi+\d\vt^a\wedge\Pi_a\,.
\end{equation}
Observe that for a given Lagrangian all the quantities 
$\,^{\rm (mat)}\E,\,^{\rm (gr)}\E_a$ and $\Omega$ 
are well-defined without any ambiguity. 

Consider the variations of the fields which vanish 
at a boundary of a region. 
We obtain the  matter field equation $\,^{\rm (mat)}\E=0$, or, explicitly, 
\begin{equation}\label{mat-6}
d\pi=(-1)^p\sigma\,,
\end{equation}
and the coframe field equation $\,^{\rm (gr)}\E_a=0$, i.e., 
\begin{equation}\label{mat-7}
d\Pi_a=-\Sigma_a\,.
\end{equation}
The left hand sides of Eq.(\ref{mat-6},\,\ref{mat-7}) are the derivatives 
of the  strengths. 
Hence, the  right hand sides of these equations 
represent the sources of the matter field and 
of the coframe (gravity) field, respectively. 

These field equations yield two conservation laws for the sources:\\
i) conservation of the matter current 
\begin{equation}\label{mat-8}
d\sigma=0\,,
\end{equation}
ii) conservation of the coframe current 
\begin{equation}\label{mat-9}
d\Sigma_a=0\,.
\end{equation}
For the generic Lagrangian used, these two conserved currents 
depend on both fields: matter field $\psi$ and 
the coframe (gravity) field $\vt^a$, i.e.,  
the currents include the contributions of two fields as well as 
the interaction between them.   
The conservation laws (\ref{mat-8},\,\ref{mat-9})  are consequences of the field 
equations. 
They, however, are strong conservation laws, because their 
 right hand sides do 
not involve the combinations of the field equations $\,^{\rm (mat)}\E$ and 
$\,^{\rm (gr)}\E_a$. This is in contrast to  the weak Noether currents that 
will appear below. 
\subsection{Noether current and charge}
In the case that the variation of the Lagrangian is an exact form,  
$\d \L=dS$, the variational 
relation  (\ref{mat-5}) again implies 
the existence of  an $(n-1)$-form $\Theta:=S-\Omega$, 
which is conserved modulo the field equations:  
\begin{equation}\label{no-1}
d \Theta+\d\psi\wedge\,^{\rm (mat)}\E+\d\vt^a\wedge\,^{\rm (gr)}\E_a=0
\quad\Longleftrightarrow \quad d \Theta\approx 0\,.
\end{equation}
Consider the variations of the field that are generated by the
Lie derivative taken with respect to a smooth vector field $\xi $, i.e.,
\begin{equation}\label{no-2}
\d \psi=L_\xi\psi=d(\xi\rfloor\psi)+\xi\rfloor d\psi\,,
\end{equation}
and 
\begin{equation}\label{no-3}
\d \vt^a=L_\xi\vt^a=d(\xi\rfloor\vt^a)+\xi\rfloor d\vt^a\,.
\end{equation}
The diffeomorphism invariance of the  $n$-form Lagrangian yields 
\begin{equation}\label{no-4}
\d \L=L_\xi\psi=d(\xi\rfloor \L)\,.
\end{equation}
Accordingly,  we have a weak (on shell) conserved odd $(n-1)$-form  
generated by the diffeomorphism symmetry of the Lagrangian 
\begin{equation}\label{no-6}
\Theta(\xi)=-\xi\rfloor \L+\Omega\,, \qquad d\Theta(\xi)\approx 0\,.
\end{equation}
We will refer to $\Theta(\xi)$ as the {\it total Noether current} of the 
matter-coframe system. 
The explicit expression for this quantity  is derived from 
(\ref{mat-5c},\,\ref{no-2},\,\ref{no-3}) as 
\begin{eqnarray}\label{no-7}
\Theta(\xi)&=&-\xi\rfloor \L+[(\xi\rfloor d\psi)+d(\xi\rfloor \psi)]\wedge \pi+
[(\xi\rfloor d\vt^a)+d(\xi\rfloor \vt^a)]\wedge \Pi_a\,.
\end{eqnarray}
This conserved current is locally constructed out of the fields appearing 
in the Lagrangian and of  the unspecified vector field $\xi$.

Extracting the total derivatives and applying the field equations, we 
decompose this current as 
\begin{equation}\label{no-8}
\Theta(\xi)=\S(\xi)+dQ(\xi)\,,
\end{equation}
where
\begin{eqnarray}\label{no-9}
\S(\xi)&=&-\xi\rfloor \L+(\xi\rfloor d\psi)\wedge \pi+(\xi\rfloor \psi)\wedge \sigma+
(\xi\rfloor d\vt^a)\wedge \Pi_a+(\xi\rfloor \vt^a)\wedge \Sigma_a\,,\nonumber\\&&
\end{eqnarray}
whereas
\begin{equation}\label{no-10}
Q(\xi):= (\xi\rfloor\vt^a)\wedge\Pi_a+(\xi\rfloor \psi)\wedge \pi\,.
\end{equation}
The currents $\Theta(\xi)$ and $\S(\xi)$ are weak conserved simultaneously. 
The current $\S(\xi)$ is algebraically linear in 
the vector field $\xi$, so  
$\S(\xi^ae_a)=\xi^a\S(e_a)$. 
Thus the cascade equations read 
\begin{eqnarray}\label{no-11}
 \xi^a&:& \qquad d\S(e_a)\approx 0\,,
\\ \label{no-12} 
d\xi^a&:& \qquad \S(e_a)\approx 0\,.
\end{eqnarray}
We rewrite them explicitly as 
\begin{eqnarray}\label{no-13}
\S(e_a)&=&-e_a\rfloor \L+(e_a\rfloor d\psi)\wedge \pi+(e_a\rfloor \psi)\wedge \sigma+
(e_a\rfloor d\vt^a)\wedge \Pi_a+\Sigma_a \approx 0\,.\nonumber\\&&
\end{eqnarray}
Thus we derive 
\begin{equation}\label{no-14}
\Sigma_a\approx e_a\rfloor \L-(e_a\rfloor d\psi)\wedge \pi-
(e_a\rfloor \psi)\wedge \sigma-
(e_a\rfloor d\vt^a)\wedge \Pi_a\,,
\end{equation}
which is the proper conserved and non-trivial energy-momentum 
current of the system.

Substituting (\ref{no-11}) into (\ref{no-8}) we obtain 
\begin{equation}\label{n-ch1}
\Theta(\xi) \approx dQ(\xi)\,,
\end{equation}
where the explicit form of the Noether charge is given in (\ref{no-10}).   
This $(n-2)$-form  is locally constructed out of the fields appearing 
in the Lagrangian and $\xi$. For a  proof that this is possible 
in  a general diffeomorphism invariant case see Ref. \cite{Lee:nz}.  
\subsection{Noether identity}
Return to the variational relation (\ref{no-1}) and consider  the case when 
the variation of the Lagrangian is an exact form. 
Eq.(\ref{no-1}) can be viewed as a condition that the equation forms $\E$ 
have to fulfill in order to yield  an exact form
\begin{equation}\label{id-1}
\d\psi\wedge\,^{\rm (mat)}\E+\d\vt^a\wedge\,^{\rm (gr)}\E_a \qquad - \,  {\rm exact}\,.
\end{equation}
In the case of diffeomorphism invariance, the first term reads 
\begin{equation}\label{id-2}
(\xi\rfloor d\psi)\wedge \,^{\rm (mat)}\E-
(-1)^p(\xi\rfloor \psi)\wedge d\sigma\, 
\end{equation}
up to a total derivative. 
Analogously, the second term in (\ref{id-1}) gives
\begin{equation}\label{id-3}
(\xi\rfloor d\vt^a)\wedge \,^{\rm (gr)}\E_a+(\xi\rfloor \vt^a)\wedge d\Sigma_a\,.
\end{equation}
The sum of the terms (\ref{id-2}) and (\ref{id-3}) should be an exact form 
for an arbitrary $\xi$. 
Observe, however, that if it is true for some vector field $\xi$ it will not be true 
for a vector field $f\xi$, where $f$ is an arbitrary function. 
The only possibility is to require the sum  of the terms 
(\ref{id-2}) and (\ref{id-3}) 
to be zero.  Thus we have  
\begin{eqnarray}\label{id-4}
(\xi\rfloor \vt^a)\wedge d\Sigma_a&=&(\xi\rfloor d\psi)\wedge \,^{\rm (mat)}\E-(-1)^p(\xi\rfloor \psi)\wedge d\sigma-\nonumber\\
&&(\xi\rfloor d\vt^a)\wedge \,^{\rm (gr)}\E_a\,.
\end{eqnarray}
On shell it means
\begin{equation}\label{id-5}
(\xi\rfloor \vt^a)\wedge d\Sigma_a\approx (-1)^{p+1}(\xi\rfloor \psi)\wedge d\sigma\,.
\end{equation}
We replace the vector field by the vector basis $\xi\to e_a$ and obtain the 
Noether identity
\begin{eqnarray}\label{id-6}
d\Sigma_a&=&(e_a\rfloor d\psi)\wedge \,^{\rm (mat)}\E-(-1)^p(e_a\rfloor \psi)\wedge d\sigma-
\nonumber\\
&&
(e_a\rfloor d\vt^b)\wedge \,^{\rm (gr)}\E_b\,,
\end{eqnarray}
or on shell
\begin{equation}\label{id-7}
d\Sigma_a\approx(-1)^{p+1}(e_a\rfloor \psi)\wedge d\sigma\,.
\end{equation}
This identity shows that, on shell, two currents, $\Sigma_a$ and $\sigma$ 
are conserved simultaneously.  

\section{A total derivative in Lagrangians}
The form (\ref{mat-1}) of the Lagrangian is not general enough  to include 
all viable Lagrangians. 
Particularly, the Hilbert-Einstein Lagrangian for gravity involves 
the second order derivatives of the metric tensor. 
The remarkable feature is that the second  derivative terms 
appear in the form of a total derivative.
We utilize this property and consider a generic  Lagrangian shifted 
by a total derivative 
\begin{equation}\label{g-inv1a}
{\widetilde \L}=\L(\psi, d\psi,\vt^a,d\vt^a)+d\Lambda(\psi, d\psi,\vt^a,d\vt^a)\,,
\end{equation}
where $\Lambda$ is an arbitrary $(n-1)$-form locally 
constructed from the fields 
and their first order derivatives only.  
The total derivative shift, as it is well known, preserves the field equations. 
Let us examine how the shift (\ref{g-inv1a}) influences  the conserved currents. 
Because the Lagrangian (\ref{g-inv1a}) involves second  derivatives 
of the dynamical fields, it is of second order due to the usual classification. 
However, because the variation operator commutes with the exterior derivative, 
the first order formalism is applicable also in this case.  
Variation of the transformed Lagrangian (\ref{g-inv1a}) takes the form  
\begin{equation}\label{g-inv1b}
\d{\widetilde \L}=\d \L+d(\d\Lambda)\,.
\end{equation}
The shift form $\Lambda$ generates additional terms, which may be collected in 
\begin{equation}\label{g-inv1b*}
\d{\widetilde \L}=\d\psi\wedge\widetilde {\sigma}+\d (d\psi)\wedge\widetilde {\pi}+\d\vt^a\wedge\widetilde 
{\Sigma}_a+\d (d\vt^a)\wedge \widetilde {\Pi}_a\,,
\end{equation}
where the shifted quantities are defined as  \cite{Hehl:ue}
\begin{eqnarray}\label{g-1}
{\widetilde \sigma}&:=&\sigma+(-1)^p d\Big(\frac{\partial\Lambda}{\partial\psi}\Big)\,,\\
\label{g-2}
{\widetilde \pi}&:=&\pi+\frac{\partial\Lambda}{\partial\psi}+(-1)^{p+1}d\Big(\frac{\partial\Lambda}{\partial(d\psi)}\Big)\,,
\end{eqnarray}
and
\begin{eqnarray}\label{g-3}
{\widetilde \Sigma}_a&:=&\Sigma_a+d\Big(\frac{\partial\Lambda}{\partial\vt^a}\Big)\,,\\
\label{g-4}
{\widetilde \Pi}_a&:=&\Pi_a-\frac{\partial\Lambda}{\partial\vt_a}-d\Big(\frac{\partial\Lambda}{\partial(d\vt_a)}\Big)\,. 
\end{eqnarray}
We extract the total derivatives in (\ref{g-inv1b*}) and obtain 
\begin{equation}\label{g-inv1bb}
\d{\widetilde \L}=\d\psi\wedge \,^{\rm (mat)}\widetilde{\E}+
\d\vt^a\wedge \,^{\rm (gr)}\widetilde{\E}_a+d(\widetilde{\Omega})\,,
\end{equation}
where the field equations forms are 
\begin{eqnarray}\label{g-5}
\,^{\rm (mat)}\widetilde{\E}&:=&{\widetilde \sigma}-(-1)^pd{\widetilde \pi}\,,\\
\,^{\rm (gr)}\widetilde{\E}_a&:=&{\widetilde \Sigma}_a+d{\widetilde \Pi}_a\,,\label{g-6}
\end{eqnarray}
and the pre-simplectic potential is 
\begin{equation}\label{g-7}
{\widetilde \Omega}:=\d\psi\wedge {\widetilde \pi}+\d\vt^a\wedge{\widetilde \Pi}_a\,.
\end{equation}
The corresponding field equations have the same form as 
(\ref{mat-6},\, \ref{mat-7}). 
Moreover, because of  (\ref{g-1}--\ref{g-4}), they are equivalent    
\begin{equation}\label{g-8}
d{\widetilde \pi}=(-1)^p{\widetilde \sigma}\quad\quad \ \Longleftrightarrow \quad 
d{ \pi}=(-1)^p{ \sigma}\,,
\end{equation}
and
\begin{equation}\label{g-9}
d{\widetilde \Pi}_a=(-1)^p{\widetilde \Sigma}_a\quad \Longleftrightarrow \quad 
d{ \Pi}_a=(-1)^p{ \Sigma}_a\,.
\end{equation}
The shifted quantities ${\widetilde \pi}$ and ${\widetilde \Pi}_a$ 
don't pick up an additional exact form. 
The conserved currents ${\widetilde \sigma}$ and ${\widetilde \Sigma}_a$  
are shifted only  by a total derivative. 
Thus we are confronted with  the known problem of an ambiguity 
of the conserved current  \cite{Wald:nt}.  
Two currents ${\Sigma}_a$ and ${\widetilde \Sigma}_a$ 
are the source terms of the field strengths. 
They also conserved simultaneously and 
yield  the same value if being integrated over a closed surface.   
Their actual values, however, are different. 

Consider the variation relation (\ref{g-inv1b}). It can be written as
\begin{equation}\label{g-10}
\d{\widetilde \L}=
\d\psi\wedge\,^{\rm (mat)}\E+\d\vt^a\wedge\,^{\rm (gr)}\E_a+d(\Omega+\d\Lambda)\,,
\end{equation}
or, on-shell,  
\begin{equation}\label{g-11}
\d{\widetilde \L}\approx d(\Omega+\d\Lambda)\,.
\end{equation}
Thus the diffeomorphism invariance generates a weak conserved current
\begin{eqnarray}\label{g-12}
{\widetilde \Theta}(\xi)&=&-\xi\rfloor {\widetilde \L} +\Omega(\xi)+
\xi\rfloor d\Lambda+
d(\xi\rfloor \Lambda)\nonumber\\
&=& -\xi\rfloor \L +\Omega(\xi)+d(\xi\rfloor \Lambda)
=\Theta(\xi)+d(\xi\rfloor \Lambda)\,.
\end{eqnarray}
Inserting the decomposition  $\Theta(\xi)=\S(\xi)+dQ(\xi)$, we obtain 
\begin{equation}\label{g-13}
{\widetilde \Theta}(\xi)=\S(\xi)+d[Q(\xi)+\xi\rfloor \Lambda]\,.
\end{equation}
Thus, the total derivative shift of the Lagrangian preserves the algebraic 
part of the Noether current, whereas it induces a shift of the Noether charge: 
\begin{equation}\label{g-14}
{\widetilde Q}(\xi)=Q(\xi)+\xi\rfloor \Lambda\,.
\end{equation}
\section{Examples}
\subsection{A $p$-form field}
Consider a Maxwell-type Lagrangian for a $p$-form field $A$  given on an 
$n$-dimensional manifold 
\begin{equation}\label{ex-3}
\L=-\frac 12 F\wedge *F\,,
\end{equation}
where $F=dA$ is the field strength and $*$ is the Hodge operator. 

Variation of the Lagrangian takes the form 
\begin{equation}\label{ex-4}
\d \L=-\frac 12 \d F\wedge *F-\frac 12 F\wedge \d*F\,.
\end{equation}
Now, we have  to calculate the variation of the form $\d*F$. 
For a well defined notion of the energy-momentum current, 
the variation $\delta\vt^a$ of the coframe field  
has to be taken into account  together with  the variation $\d A$ 
of the $p$-form field. 
In this case the variational operator does not commute  
with the Hodge operator, $\d*\ne *\d$. 
We will use here the master formula \cite{Muench:1998ay}, 
which, in the case of (pseudo-)orthonormal coframe, takes the form
\begin{equation}\label{ex-5}
(\d*-*\d)F=\d\vt^a\wedge (e_a\rfloor *F)-*\left[\d\vt^a\wedge(e_a\rfloor F)\right]\,.
\end{equation}
Hence, 
\begin{eqnarray}\label{ex-6}
F\wedge \d*F&=&F\wedge *\d F+\d\vt^a\wedge \Big[(-1)^{p+1}F\wedge(e_a\rfloor *F)-
(e_a\rfloor F)\wedge *F\Big]\,,
\end{eqnarray}
and (\ref{ex-4}) reads,   
\begin{equation}\label{ex-7}
\d \L=\d F\wedge *F+\frac 12 \d\vt^a\wedge 
\Big((e_a\rfloor F)\wedge *F+(-1)^{p}F\wedge(e_a\rfloor *F)\Big)\,.
\end{equation}
Thus, the field momentum  (\ref{form-2}) takes the value $\pi= *F$, 
while the  current of the field $A$ vanishes,  $\sigma=0$. 
Hence, the field equation reads 
\begin{equation}\label{ex-7x}
d*F=0\,.
\end{equation}
Using (\ref{ex-7}), we obtain the  energy-momentum current as 
 \begin{equation}\label{ex-8}
 \Sigma_a=\frac 12 \Big((e_a\rfloor F)\wedge *F+
(-1)^pF\wedge(e_a\rfloor *F)\Big)\,,
\end{equation}
and the Noether charge as   
\begin{equation}\label{ex-9}
 Q(e_a)=(e_a\rfloor A)\wedge *F\,.
\end{equation}
For the Maxwell field ( $p=1$) on a 4D-manifold, Eq.(\ref{ex-8}) gives the correct result.
 
The trace of the energy-momentum tensor, corresponding to the 
current (\ref{ex-8}), is proportional to the $n$-form $\vt^a\wedge  \Sigma_a$. 
Calculate 
 \begin{equation}\label{ex-10}
\vt^a\wedge \Sigma_a=
-\frac 12(n-2p-2)F\wedge *F\,.
\end{equation}
Thus the corresponding energy-momentum tensor is traceless if and only if 
$n=2(p+1)$, i.e., in the case when the strength $F$ is a middle form 
(for even $n$). 

The antisymmetry part of the energy-momentum tensor  is proportional to the 
$(n-2)$-form  $e^a\rfloor \Sigma_a$, see, for instance, \cite{Itin:2001xz}. 
Calculating this expression, we obtain 
\begin{equation}\label{ex-10b}
e^a\rfloor \Sigma_a=0\,.
\end{equation}
Thus, the energy-momentum tensor of the $p$-form field is symmetric for 
an arbitrary value of the degree $p$.
\subsection{Vacuum GR in coframe representation}
Let  a differential  manifold $M$ be endowed with a 
{\it pseudo-orthonormal coframe} 
field $\vt^a$. It means that the metric on $M$ can be represented as 
\begin{equation}\label{ex-1}
g=\eta_{ab}\vt^a\otimes\vt^b\,,
\end{equation}
where $\eta_{ab}=diag(-1,1,\cdots,1)$. 
Consequently the Hodge map, which depends on the metric tensor, 
acts on the basis forms as follows, 
\begin{equation}\label{ex-2}
*(\vt^{a_1}\wedge\cdots\wedge\vt^{a_q})=\epsilon^{a_1\cdots a_q\cdots a_n}
\vt_{a_{q+1}}\wedge\cdots\wedge\vt_{a_n}\,,
\end{equation}
where the indices are lowered accordingly to $\vt_a:=\eta_{ab}\vt^b$.  

The Einstein-Hilbert Lagrangian  
corresponds to the coframe Lagrangian \cite{Rumpf}, \cite{Mielke} as 
\begin{eqnarray}\label{ex-10c}
\L&=&\frac 12 \Big[(\C_a\wedge\vt^b)\wedge*(\C_b\wedge\vt^a)
-2(\C_a\wedge\vt^a)\wedge*(\C_b\wedge\vt^b)\Big]+d\Lambda\,,
\end{eqnarray}
where $\C^a:=d\vt^a$ and 
\begin{equation}\label{ex-10a}
\Lambda=\vt^a\wedge*\C_a\,.
\end{equation}
Using the notation
\begin{equation}\label{ex-11}
\F^a:=\C^a-2e^a\rfloor(\vt^m\wedge \C_m)-\vt^a\wedge (e_m\rfloor \C^m)\,,
\end{equation}
the Lagrangian can be written in a compact form as 
\begin{equation}\label{ex-12}
\L=\frac 12 \C_a\wedge*\F^a+d\Lambda\,.
\end{equation}
Its variation  can be written \cite{Itin:2001bp,Itin:2001xz} as 
 \begin{equation}\label{ex-13}
\d \L=\d C^a\wedge*\F_a+\d\vt^a\wedge \Big(e_a\rfloor \L-(e_a\rfloor \C^b)\wedge *\F_b\Big)
+d(\d \Lambda)\,.
\end{equation}
Thus we identify the field momentum and the conserved current, respectively
\begin{equation}\label{ex-14}
\Pi^a= *\F_a\,,\qquad \Sigma^a=e_a\rfloor \L-(e_a\rfloor \C^b)\wedge *\F_b\,.
\end{equation}
Consequently the field equation is 
\begin{equation}\label{ex-16}
d *\F^a=\Sigma^a\,.
\end{equation}
The canonical Noether charge is shifted accordingly to 
\begin{equation}\label{ex-17}
Q(e_a)=*\F_a+e_a\rfloor \Lambda \,.
\end{equation}
The Lagrangian (\ref{ex-10}) is invariant under two different 
groups of symmetries:  

i) The (pseudo-)group of diffeomorphism transformations of the 
manifold which equivalent to the set of coordinate transformations.    
Such an invariance is usually referred to as  general covariance. 
All the quantities introduced above 
(including the conserved current!)  are 
manifestly invariant under these transformations. 

ii) The group of local (pointwise) transformations of the coframe. 
The Lagrangian preserves its form if we replace the coframe by 
\begin{equation}\label{ex-18}
\vt^a\to{A^a}_b(x)\vt^b, \quad {\textrm {where for all  } }\, x \quad {A^a}_b(x)\in SO(1,3)\,.
\end{equation}

Due to  the well known theorem of the calculus of  variations, 
the field equation (\ref{ex-16}) is 
invariant under these transformations. 
However, the separation of this equation to the exact form in the 
left hand side and 
 the conserved current on the right hand side is not invariant. 
Thus also the diffeomorphism invariant conserved current $\Sigma_a$ 
is not invariant 
under the local ``internal'' transformations (\ref{ex-18}). 
Certainly, this result  corresponds to the known fact that every 
covariant expression constructed from the first order derivatives the 
of metric is  trivial. 
Actually, in the view of the complete group of invariance transformation  
of the Lagrangian,  
the conserved current (\ref{ex-14}) is only a type of a pseudo-tensor. 

\subsection{Metric-free electrodynamics}
In the axiomatic approach to classical electrodynamics 
\cite{gentle,book,ROH2002}, spacetime is considered as a 
4-dimensional differentiable manifold without any additional geometrical 
structure (metric or connection).

The {\sl first axiom} of the  electric charge conservation $dJ=0$ yields 
the field equation 
\begin{equation}\label{ax-1}
dJ=0\qquad \Longrightarrow \qquad dH=J\,.
\end{equation}

The {\sl second axiom} postulates 
the existence of the {Lorentz force density}
\begin{equation}\label{ax-2}
f_\alpha=(e_\alpha\rfloor F)\wedge J\,.
\end{equation}

The {\sl third axiom} requires the magnetic flux conservation
\begin{equation}\label{ax-3}
dF=0\qquad \Longrightarrow \qquad F=dA\,.
\end{equation}

Here $J$ is the electric current density 3-form, 
$F=(E,B)$ is the untwisted  2-form of the electromagnetic field strength, 
$H=({\cal H},{\cal D})$ is the twisted 2-form of the  
electromagnetic excitation, and 
$e_a$ is a  frame field on the manifold. 
A fairly detailed account, including the conventions and references to the
literature, can be found in \cite{book}.

To complete the formulation, a relation between $H$ and $F$ 
is required, namely the constitutive law. 
This constitutive law is postulated to be  {\em local and linear} ("linear
electrodynamics"). 
The relation between the two forms is established by an odd 
{\it constitutive tensor} $\kappa$, which maps even 2-forms to 
odd 2-forms and vice versa.
Namely, $H=\kappa( F)$. 
We want to find a   componentwise representation of  $\kappa $. 
Introduce a coframe field $\vt^a$, which is dual to the frame field $e_a$ 
appearing in (\ref{ax-2}). Thus, $\vt^{ab}= \vt^a\wedge\vt^b$ is a basis for even 2-forms. 
Using the Levi-Civita pseudo-tensor $\epsilon_{abcd}$
 a volume element $\epsilon$ can be  defined by the coframe as 
$\epsilon= \epsilon_{abcd}\vt^{abcd}/4!$. 
Thus,  $\epsilon_{ab}=e_a\rfloor e_b\rfloor \epsilon $ is the basis for odd 2-forms. 
Accordingly, in the components,  
\begin{equation}\label{ex-18a}
F=\frac 12 F_{ab}\vt^{ab}, \qquad
H=\frac 12 H^{ab}\epsilon_{ab}\,.
\end{equation}
Linearity of the $\kappa$-map means $\epsilon_{ab}=\kappa(\vt^{ab})$. 
Thus, the tensor representation $\chi^{mnab}$ of the operator $\kappa$ 
reads  
\begin{equation}\label{axiom5a}
\kappa(\vt^{ab}):=\frac 12 \chi^{mnab}\ep_{mn}\, 
\end{equation}
and
\begin{equation}\label{axiom5b}
H=\kappa (F)=\frac 12 F_{ab}\,\kappa(\vt^{ab})=\frac 14F_{ab}\, \chi^{mnab}\ep_{mn} \,,
\end{equation}
or, in  components,  
\begin{equation}\label{axiom5d}
H^{mn}=\frac 12\chi^{mnab}F_{ab}\,.
\end{equation}

Consider the ``ordinary'' Lagrangian  of the free electromagnetic field 
\begin{equation}\label{lagr1}
\L=-\frac 12 F\wedge H=-\frac 12 F\wedge \kappa(F)\,.
\end{equation} 
Rewrite the Lagrangian componentwise as 
\begin{equation}\label{lagr3}
\L=-\frac 1{16}F_{ab}\,\vt^{ab}\wedge F_{pq}\,\chi^{pqmn}\,\ep_{mn}=
\frac 18 F_{ab}F_{cd}\,\chi^{abcd}\ep\,.
\end{equation} 

Under the action of the group $GL(4,\mathbb R)$ the tensor $\chi^{abcd}$ 
can be  irreducibly decomposed into 3 pieces, 
\begin{equation}\label{axiom5e}
\chi^{abcd}=\,^{(1)}\!\chi^{abcd}+\,^{(2)}\!\chi^{abcd}+\,^{(3)}\!\chi^{abcd}\,,
\end{equation}
where
\begin{eqnarray}\label{ax8x}
{}^{(3)}\!\chi^{abcd}&:=&\chi^{[abcd]}\,,\\
\label{ax9}
{}^{(2)}\!\chi^{abcd}&:=&
\frac 12 \left(\chi^{abcd}-\chi^{cdab}\right)\,,\\
\label{ax10}
{}^{(1)}\!\chi^{abcd}&:=&\chi^{abcd}-{}^{(2)}\chi^{abcd}-{}^{(3)}\chi^{abcd}\,.
\end{eqnarray}
The Lagrangian (\ref{lagr3}) has an additional symmetry: 
\begin{equation}\label{symm2}
 \chi^{abcd}= \chi^{cdab}\,.
\end{equation}
Thus the {\it skewon part} ${}^{(2)}\chi^{\alpha\beta\gamma\delta}$ 
does not contribute to this Lagrangian. 

We are looking for the energy-momentum current of the electromagnetic fields 
$F$ and $H$. 
Due to the linearity of $\kappa$, the constitutive tensor depends only 
on the non-electromagnetic 
(geometric) variables. So, it variation will not give a contribution to the 
energy-momentum current of the electromagnetic field $\Sigma_a$. 
We will show now that, in order to obtain the true expression for $\Sigma_a$, 
it is enough to take into account only the variation of the potential $A$ 
and of the coframe $\vt^a$.

We have 
\begin{equation}\label{lagr2}
\delta\L=-\frac 12 \d(F\wedge \k F)\,.
\end{equation} 
Using  the symmetry (\ref{symm2}), we rewrite the right hand side as 
\begin{eqnarray}\label{lagr6}
\d\L&=&\frac 1{4}\d(F_{ab})\,H^{ab}\ep
+\frac 1{8}F_{ab}\,H^{ab}\,\d(\ep)\,.
\end{eqnarray} 
Observe now, that the variations in right hand side of (\ref{lagr6}) 
do not depend on the odd operator used. 
Thus, analogously to the standard Maxwell theory and to the $p$-form 
field expression (\ref{ex-7}), we may write 
\begin{equation}\label{lagr7}
\d \L=\d F\wedge H
+ \d\vt^a\wedge \Sigma_a\,.
\end{equation}
where $\Sigma_a=\partial \L/\partial \vt^a$ is 
 the Hilbert energy-momentum current:  
\begin{equation}\label{lagr8}
\Sigma_a=\frac 12 \left[(e_a\rfloor F)\wedge H-F\wedge(e_a\rfloor H)\right]\,.
\end{equation}
Using the commutativity of the operators $\d$ and $d$ and extracting 
the total derivative,   
we obtain the variation of the Lagrangian (\ref{lagr1}) in the form
\begin{equation}\label{lagr9}
\delta\L=d(\d A\wedge H)+\d A\wedge dH+\d\vt^a\wedge\Sigma_a\,.
\end{equation} 
In accordance with the non-dynamical coframe procedure (see section 2),  
we extract from this relation only one field equation $dH=0$, which is 
in accordance with  (\ref{ax-1}) providing $J=0$.
For the field that satisfies this field equation, the variation (\ref{lagr9}) reads  
\begin{equation}\label{lagr11}
\delta\L=d(\d A\wedge H)+\d\vt^a\wedge\Sigma_a\,.
\end{equation}
Consider now  the variations generated by diffeomorphism invariance 
of the Lagrangian: 
\begin{eqnarray}
\d A&=&\xi\rfloor d A+d(\xi\rfloor A)\,,\\
\d\vt^a&=&\xi\rfloor d \vt^a+d(\xi\rfloor\vt^a)\,,\\
\delta\L&=&d(\xi\rfloor \L)\,. 
\end{eqnarray}
We take for non-dynamical coframe $\d\vt^a=0$. Consequently,  the relation 
(\ref{lagr11}) becomes 
\begin{equation}\label{lagr12}
d\Theta(\xi)+d(\xi\rfloor\vt^a)\wedge\Sigma_a=0\,,
\end{equation}
where 
\begin{equation}\label{lagr13}
\Theta(\xi)=-\xi\rfloor \L+[\xi\rfloor d A+d(\xi\rfloor A)]\wedge H\,.
\end{equation}
Extracting the total derivative, we may write it as 
\begin{equation}\label{lagr14}
\Theta(\xi)=\S(\xi)+dQ(\xi)\,
\end{equation}
with 
\begin{eqnarray}\label{lagr15}
Q(\xi)&=&\xi\rfloor A\wedge H\,,\\
\label{lagr16}
\S(\xi)&=&-\xi\rfloor \L+\xi\rfloor d A\wedge H-\xi\rfloor A\wedge dH\,.
\end{eqnarray}
On shell $dH=0$. Thus  (\ref{lagr12}) becomes  
\begin{equation}\label{lagr17}
d\S(\xi)+d(\xi\rfloor\vt^a)\wedge\Sigma_a=0\,,
\end{equation}
with 
\begin{equation}\label{lagr19}
\S(\xi)=-\xi\rfloor \L+\xi\rfloor F\wedge H\,.
\end{equation}
The first cascade equation (\ref{nd-9}) yields the conservation law 
\begin{equation}\label{lagr20}
d\S(e_a)=0\,,
\end{equation}
whereas  the second  cascade equation turns out to be an identity
$\S(e_a)=-\Sigma_a$. 
Thus we have derived the conservation law for the Hilbert current 
\begin{equation}\label{lagr21}
\Sigma_a=e_a\rfloor \L-e_a\rfloor F\wedge H\,, 
 \end{equation}
which is shown to be associated with the diffeomorphism invariance of 
the Lagrangian.
\section{Outlook of main results}
Let us summarize the results of our consideration. 
We have examined the Noether procedure for a diffeomorphism 
invariant Lagrangian in three different ways: 
\begin{itemize}
\item[1.] We considered the variation of the Lagrangian, 
taking  into account only the field 
itself and not involving any information on the geometry of the manifold. 
In this way, we have obtained a conserved current which identically 
vanishes on shell.  
\item[2.] The second consideration was based on a non-dynamical coframe. 
Corresponding cascade equation turned out to describe the equivalence 
between canonical and coframe (Hilbert) current.
In this way, we derived a conserved current which, on shell, 
does not vanish identically.  

\item[3.] We have considered a system of two dynamical fields: a $p$-form field 
$\psi$ and a coframe field $\vt^a$.  
The Noether procedure is consistent in this case. 
The vanishing canonical current of the system 
represents a relation between the Noether 
and the Hilbert currents. 
Consequently, the conservation law for the canonical current of the $p$-form field 
yields the conservation law for the Hilbert currents of this field.
In this way the Hilbert current is related to the diffeomorphism invariance of 
the Lagrangian. Consequently it obtains the status of {\it the} 
energy-momentum current. 
Moreover, the Hilbert current of the system is shown 
 to be the source of the coframe field. 
We derive also an expression for the Noether charge.
\end{itemize}
The main result of our consideration is: In order to have a complete and 
non contradictory Noether current, it is necessary to involve the geometry of 
the manifold, such as the coframe field, in the variational procedure. 

This result may be correlated with Dicke's analysis \cite{dicke}  which shows 
that the gravity field is unique in having a Lagrangian describing an 
``interaction only with itself''. 
All other viable Lagrangians have to involve the metric. 
Thus, in fact, they  describe an interaction with  the gravitational field. 
\section*{Acknowledgments}
I am grateful to S. Kaniel and G. Kalai  
for constant support and to F.~W.~Hehl for viable suggestions. 
My gratitude to anonymous  referees for their active interest and 
important comments. 
\section*{References}


\begin{thebibliography}{99}


\bibitem{Julia:1998ys}
B.~Julia and S.~Silva,
Class.\ Quant.\ Grav.\  {\bf 15}, 2173 (1998)
[arXiv:gr-qc/9804029].

\bibitem{Julia:2000er}
B.~Julia and S.~Silva,
Class.\ Quant.\ Grav.\  {\bf 17}, 4733 (2000)
[arXiv:gr-qc/0005127].

\bibitem{Wald:1999wa}
R.~M.~Wald and A.~Zoupas,
Phys.\ Rev.\ D {\bf 61}, 084027 (2000)
[arXiv:gr-qc/9911095].


\bibitem{Chen:1998aw}
C.~M.~Chen and J.~M.~Nester,
Class.\ Quant.\ Grav.\  {\bf 16}, 1279 (1999)
[arXiv:gr-qc/9809020].

\bibitem{Nester:du}
J.~M.~Nester and R.~S.~Tung,
Phys.\ Rev.\ D {\bf 49}, 3958 (1994)
[arXiv:gr-qc/9401002].

\bibitem{Fatibene:2000jm}
L.~Fatibene, M.~Ferraris, M.~Francaviglia and M.~Raiteri,
J.\ Math.\ Phys.\  {\bf 42}, 1173 (2001)
[arXiv:gr-qc/0003019].

\bibitem{Fatibene:nc}
L.~Fatibene, M.~Ferraris and M.~Francaviglia,
J.\ Math.\ Phys.\  {\bf 38}, 3953 (1997).


\bibitem{Iyer:1995kg}
V.~Iyer and R.~M.~Wald,
Phys.\ Rev.\ D {\bf 52}, 4430 (1995)
[arXiv:gr-qc/9503052].


\bibitem{Lee:nz}
J.~Lee and R.~M.~Wald,
J.\ Math.\ Phys.\  {\bf 31} (1990) 725.

\bibitem{Gotay:1997eg}
M.~J.~Gotay, J.~Isenberg and J.~E.~Marsden,
arXiv:physics /9801019.

\bibitem{Gotay} M.~J.~Gotay, J.~E.~Marsden,
Contemp. Math., {\bf 132}, 
Amer. Math. Soc., Providence, RI, (1992). 


\bibitem{Trautman} A.~Trautman, 
Acta Phys.\ Polon.\ B {\bf 27}, 839 (1996)

\bibitem{Sardanashvily:hf}
G.~Sardanashvily,
Class.\ Quant.\ Grav.\  {\bf 14}, 1357 (1997).


\bibitem{Bak:1993us}
D.~Bak, D.~Cangemi and R.~Jackiw,
Phys.\ Rev.\ D {\bf 49}, 5173 (1994)
[Erratum-ibid.\ D {\bf 52}, 3753 (1995)]
[arXiv:hep-th/9310025].



\bibitem{Brown:1992br}
J.~D.~Brown and J.~W.~York,
Phys.\ Rev.\ D {\bf 47}, 1407 (1993).

\bibitem{Anderson:1996sc}
I.~M.~Anderson and C.~G.~Torre,
Phys.\ Rev.\ Lett.\  {\bf 77}, 4109 (1996)
[arXiv:hep-th/9608008].

\bibitem{Torre:1997cd}
C.~G.~Torre,
arXiv:hep-th/9706092.



\bibitem{Szabados:1991dd}
L.~B.~Szabados,
Class.\ Quant.\ Grav.\  {\bf 9}, 2521 (1992).

 \bibitem{Dubois-Violette:1986ws}
M.~Dubois-Violette and J.~Madore,
Comm.\ Math.\ Phys.\  {\bf 108}, 213 (1987).

\bibitem{Hehl:ue}
F.~W.~Hehl, J.~D.~McCrea, E.~W.~Mielke and Y.~Neeman,
Phys.\ Rept.\  {\bf 258}, 1 (1995)
[arXiv:gr-qc/9402012].

\bibitem{Wald:nt}
R.~M.~Wald,
Phys.\ Rev.\ D {\bf 48} (1993) 3427
[arXiv:gr-qc/9307038].

\bibitem{Iyer:1994ys}
V.~Iyer and R.~M.~Wald,
Phys.\ Rev.\ D {\bf 50} (1994) 846
[arXiv:gr-qc/9403028].

\bibitem{Henneaux:1996ws}
M.~Henneaux, B.~Knaepen and C.~Schomblond,
Comm.\ Math.\ Phys.\  {\bf 186}, 137 (1997)
[arXiv:hep-th/9606181].

\bibitem{Freedman:us}
D.~Z.~Freedman and P.~K.~Townsend,
Nucl.\ Phys.\ B {\bf 177}, 282 (1981).

\bibitem{Muench:1998ay}
U.~Muench, F.~Gronwald and F.~W.~Hehl,
Gen.\ Rel.\ Grav.\  {\bf 30} (1998) 933
[arXiv:gr-qc/9801036].

\bibitem{dicke}R. H. Dicke, \textit{The Theoretical
Significance of Experimental Relativity. Appendix 4} Gordon and Breach,
New York (1965).

\bibitem{Rumpf} H.~Rumpf, 
\emph{Z. Naturf.} \textbf{33a} (1978) 1224--1225.

\bibitem{Kopczynski:af}
W.~Kopczynski,
Annals Phys.\  {\bf 203}, 308 (1990).

\bibitem{Mielke}
E.~W.~Mielke,
Annals Phys. (NY)  {\bf 209}, 78 (1992).

\bibitem{Itin:2001bp}
Y.~Itin,
Class.\ Quant.\ Grav.\  {\bf 19} (2002) 173
[arXiv:gr-qc/0111036].

\bibitem{Itin:2001xz}
Y.~Itin,
Gen.\ Rel.\ Grav.\ {\bf 34} (2002) 1819 
[arXiv:gr-qc/0111087]. 



\bibitem{gentle} 
F.~W.~Hehl and Yu.~N.~Obukhov, 
arXiv:physics/0005084.

\bibitem{book} F.~W.~Hehl and Yu.~N.~Obukhov, {\it Foundations of
Classical Electrodynamics\/}, Birkh\"auser, Boston, MA (2003) to be
published.



\bibitem{ROH2002} G.F.\ Rubilar, Yu.N.\ Obukhov and F.W.\ Hehl, 
{ Int.\ J.\ Mod.\ Phys.} {\bf D11} (2002) No.8 or 9.

\end{thebibliography}
\end{document}